\documentclass[letter,twocolumn]{jpsj3}
\usepackage{txfonts}
\usepackage{cite}
\usepackage{braket}
\usepackage{bm}
\usepackage{amsmath}

\usepackage{color}
\usepackage{ulem}

\title{Time-reversal symmetry breaking superconductivity in hole-doped monolayer MoS$_2$}

\author{Rikuto Oiwa$^1$, Yuki Yanagi$^2$, and Hiroaki Kusunose$^1$}
\inst{$^1$Department of Physics, Meiji University, Kawasaki 214-8571, Japan \\
$^2$Institute for Materials Research, Tohoku University, Sendai, Miyagi 980-8577, Japan} 

\abst{
We investigate the nature of the time-reversal breaking pairing state in the hole-doped monolayer MoS$_2$ on the basis of the realistic three-orbital attractive Hubbard-like model with the atomic spin-orbit coupling. 
Due to the multi-band features arising from the Mo $d$ orbitals in the noncentrosymmetric crystal structure, the Lifshitz transition takes place upon hole doping. Across the Lifshitz transition point, the sign of the relative phase between the Cooper-pair components drastically changes, leading to the emergence of the time-reversal breaking phase with complex gap functions.
It is shown that this intriguing pairing state is characterized by the finite momentum-space distributions of the orbital and spin angular momentum with three-fold rotational symmetry on the Fermi-surface pockets around K and K$'$ points.
The present mechanism for the time-reversal breaking superconductivity can ubiquitously be applied to spin-orbit-coupled metals in noncentrosymmetric crystal structures.
}

\begin{document}
\maketitle

\textit{Introduction |}
Superconductivity is characterized by broken U(1) gauge symmetry. 
In addition to that, the so-called unconventional superconductors also spontaneously break crystalline and/or time-reversal symmetries.
In particular, superconductivity that breaks time-reversal symmetry (TRS) has attracted continuous attention for their intriguing features.
Although nonunitary spin-triplet pairing states have been discussed extensively in the representative compounds such as UPt$_{3}$~\cite{H.Tou, R.Joynt, C.Pfleiderer, Y.Machida} and Sr$_{2}$RuO$_{4}$~\cite{A.P.Mackenzie, C.W.Hicks, J.Xia, A.Kapitulnik, S.Raghu,Maeno,Yanase}, no concrete evidences for spin-singlet TRS breaking states have been observed.
Nevertheless, a possibility of spin-singlet TRS breaking states is examined recently in several candidate materials, such as doped graphene~\cite{M.L.Kiesel(graphene), R.Nandkishore}, water-intercalated Na$_{x}$CoO$_{2}\cdot$$y$H$_{2}$O~\cite{Baskaran,Ogata,M.L.Kiesel(Co)} and Fe-based superconductors~\cite{A.D.Christianson, J.Garaud, Z.L.Mahyari, J.Garaud(2), Y.Gao, V.Stanev, M.Silaev, M.Marciani, T.T.Ong, R.M.Fernandes, F.Ahn, S.Maiti, J.Garaud(3)}.
These TRS breaking superconductivities are commonly characterized by having multi-band components, which have a potential ``frustration'' in their relative phases.
Such a frustration can lead to an intermediate relative phase between $0$ and $\pi$, exhibiting a TRS breaking state
even in $s$-wave symmetry~\cite{V.Stanev, G.Blumberg, Y.Tanaka, J.Carlstrom, C.Platt, Agterberg, X.Hu, S.-Z.Lin, S.Maiti(2),Y.Tanaka(2),Y.Yerin}
.

In recent years, the layered transition-metal dichalcogenides (TMDCs), $MX_{2}$ ($M$=Mo, W and $X$=S, Se, and Te) have extensively been investigated as a new platform for exotic superconductivity in context of noncentrosymmetric spin-orbit physics.
In particular, an ion-gated (\textit{electron-doped}) MoS$_{2}$ exhibits peculiar superconductivity with remarkably large upper critical field~\cite{Iwasa}. 
Moreover, the exotic topological spin-singlet $p + i p$ state is discussed theoretically, which has nonzero Chern numbers and spontaneously breaks TRS~\cite{N.F.Q.Yuan}.

In addition to these fascinating properties, the \textit{hole-doped} TMDCs are more interesting to have a richer variety of pairing states than that of the electron doping due to the multi-band features in the valence bands originating from the transition-metal $d$ orbitals with relatively strong spin-orbit coupling (SOC)~\cite{Y.-T.Hsu, E.Sosenko, W.-Y.He}.
It is proposed that a mixed spin-singlet and triplet state is realized, which is robust against large in-plane magnetic fields, at very low doping in monolayer MoS$_{2}$~\cite{E.Sosenko} and NbSe$_{2}$, where the latter can be viewed as heavily hole-doped MoS$_{2}$~\cite{D.Mockli}.
It is also proposed that the topological superconducting state with the bulk nodal points along the $\Gamma-$M lines can be driven by the onsite attraction in a family of TMDCs, NbSe$_2$ and TaS$_2$, under the magnetic field~\cite{W.-Y.He}. 
The present authors also discuss the possible pairing states in the hole-doped monolayer MoS$_2$ at various hole doping by solving the linearized BCS gap equations based on the realistic tight-binding model with SOC and Hubbard-like attractions~\cite{R.Oiwa}.

In this letter, we further investigate the nature of the superconducting states in the hole-doped monolayer MoS$_2$ in whole temperature range at various hole dopings.
We found the additional TRS breaking pairing state inside the TRS preserving superconducting phase in the vicinity of the Lifshitz transition.
This $s$-wave TRS breaking state can emerge only in the presence of SOC, because multiple gap components can hybridize with each other only in the presence of the SOC due to symmetry.
Moreover, the Lifshitz transition itself is realized with the spin-split bands due to the SOC in the noncentrosymmetric crystal structure.
These are essential ingredients for the occurrence of the $s$-wave TRS breaking state in this system.
The TRS breaking state is characterized by the symmetric momentum-space distributions of the spin and orbital angular momentum with three-fold rotational symmetry on the Fermi surfaces (FS) around the K and K$'$ points.

\textit{Three-orbital BCS Hamiltonian |} 
Let us begin with the realistic tight-binding model Hamiltonian introduced in Refs. \citen{G.-B.Liu} and \citen{R.Oiwa},
\begin{align}
&
H_{0}=\sum_{\bm{k}}\sum_{\sigma}^{\uparrow,\downarrow}\sum_{mm'}^{0,\pm2}c_{\bm{k}m\sigma}^{\dagger}\left[\mathcal{H}_{0}(\bm{k})\right]_{mm'}^{\sigma\sigma}c_{\bm{k}m'\sigma}^{},
\cr&\quad
\left[\mathcal{H}_{0}(\bm{k})\right]_{mm'}^{\sigma\sigma}=\varepsilon_{mm'}(\bm{k})-\mu\,\delta_{mm'}
+\frac{\lambda}{2}m\sigma\,\delta_{mm'},
\label{eq:ham0}
\end{align}
where $m$ and $\sigma$ represent the partially occupied Mo $d$ orbitals, $\ket{0} \equiv \ket{d_{z^{2}}}$, $\ket{\pm 2} \equiv \left(\,\ket{d_{x^{2}-y^{2}}}\pm i \ket{d_{xy}}\,\right)/\sqrt{2}$, and the $z$ component of the spin, respectively.
The latter is a good quantum number because the SOC is the Ising type.
The strength of the SOC ($\lambda\sim 0.073$ eV) and 
hopping parameters in the kinetic energy $\varepsilon_{mm'}(\bm{k})$, measured from the chemical potential $\mu$, are taken from the first-principles band calculation~\cite{G.-B.Liu}.

\begin{figure}[t!]
\begin{center}
\includegraphics[width=7.0cm]{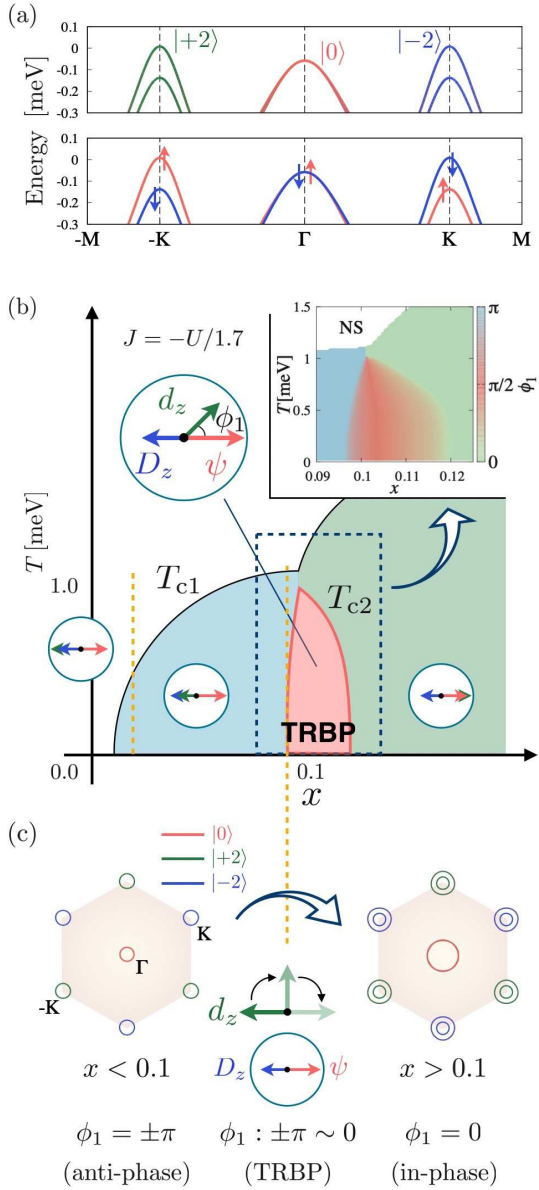}
\vspace{2mm}
\caption{
Valence band structures, the doping vs. temperature phase diagram, and the change of the FS topology as doping in the monolayer MoS$_2$.
(a) The orbital and spin dependences of the valence-band dispersions. 
(b) The superconducting phase diagram for $U=-0.5$, $J =-U/1.7$ eV.
The time-reversal symmetry breaking phase (TRBP) with the complex spin-triplet gap function $d_{z}$ appears in the vicinity of the Lifshitz transition.
The color map in the inset represents the relative phase $\phi_{1}$ between $d_{z}$ and the spin-singlet component $\psi$.
(c) The change of the Fermi-surface topology (Lifshitz transition) and the corresponding relative phase $\phi_{1}$ near $x \sim 0.1$.
}
\label{fig:1}
\end{center}
\vspace{-7mm}
\end{figure}

The obtained energy dispersions near the valance band edge are shown in Fig.~\ref{fig:1}(a) with the orbital and spin dependences in the upper and lower panels, respectively. 
In the vicinity of the valance band edge, the spin-split energy bands around K and K$'$ pockets consist of the $d_{x^{2}-y^{2}}$ and $d_{xy}$ orbitals, while the almost degenerate bands originating from $d_{z^{2}}$ orbital appear around $\Gamma$ pocket.
Note that the spin splittings are caused by the SOC in the noncentrosymmetric crystal structure, and the magnitude of the splitting is roughly given by $2\lambda$.

\textit{Gap equaitons |}
Since the superconductivity realized in the ion-gated electron doping is most likely as an $s$-wave pairing~\cite{Iwasa}, we introduced the isotropic effective direct and exchange interactions in the previous study~\cite{R.Oiwa} in order to discuss the possible pairing states in the hole-doped case as well, in which $U$ is the direct attractive interaction, while $J_{0}$ and $J_{2}$ are the inter- and intra-orbital exchange interactions between $\ket{0}$ and $\ket{\pm2}$, respectively.
We have set $J_{0}=J_{2}$ for simplicity.
According to the point group symmetry of $D_{3h}$~\cite{R.Oiwa}, the gap function is classified by A$'_{1}$, E$'$, and E$''$ irreducible representations.
Among them A$'_{1}$ state has much higher $T_{\rm c}$ than the others, and hence we restrict our discussion to A$'_{1}$ pairing state.
The gap function in the A$'_{1}$ irreducible representation is given in the form,
\begin{align}
&\Delta_{\rm s}(\phi)
=\frac{1}{\sqrt{2}}\psi X_{1}+|D_{z}|e^{i\phi_{2}}X_{2}+|d_{z}|e^{i\phi_{1}}X_{3},
\cr&\quad
\bm{X}=\left[\,\tau^{0}(i\sigma_{y}),\,\, (i\tau^{y})(i \sigma_{z} \sigma_{y}),\,\,(i \tau^{z}\tau^{y})(i\sigma_{y})\,\,\right],
\label{eq:gap_A1}
\end{align}
where $\bm{\sigma}$ is the Pauli matrices for spin space, while the matrices in the orbital space ($\ket{0},\ket{+2}$, $\ket{-2}$) are given as
\begin{align}
&
\tau^{x}=\begin{pmatrix} 0 & 0 & 0 \\ 0 & 0 & 1 \\ 0 & 1 & 0 \end{pmatrix},
\,\,\,
\tau^{y}=\begin{pmatrix} 0 & 0 & 0 \\ 0 & 0 &-i \\ 0 & i & 0 \end{pmatrix},
\cr&
\tau^{z}=\begin{pmatrix} 0 & 0 & 0 \\ 0 & 1 & 0 \\ 0 & 0 &-1 \end{pmatrix},
\,\,\,
\tau^{0}=\begin{pmatrix} \sqrt{2} & 0 & 0 \\ 0 & 0 & 0 \\ 0 & 0 & 0 \end{pmatrix}.
\end{align}
The conventional spin-singlet (SS) pairing near $\Gamma$ pocket is described by $\psi$, while 
the spin-triplet orbital-singlet (ST-OS) [spin-singlet orbital-triplet (SS-OT)] pairing near K and K$'$ pockets is denoted by $d_{z}$ [$D_{z}$].
Note that the predominant $\psi$ can always be taken as real, while $d_{z}$ and $D_{z}$ can be complex with the relative phases, $\phi=(\phi_{1}$, $\phi_{2})$.

The matrix representation of the Bogoliubov-de Gennes (BdG) Hamiltonian at $\bm{k}$ is given by
\begin{align}
\mathcal{H}_{\rm BdG} (\bm{k}; \phi) = 
\left(
\begin{matrix}
\mathcal{H}_{0} (\bm{k}) & \Delta_{\rm s}(\phi) \\
\Delta^{\dagger}_{\rm s}(\phi) & -\mathcal{H}^{*}_{0} (-\bm{k})
\end{matrix}
\right),
\label{eq:BdG}
\end{align}
where the matrix elements of $\mathcal{H}_{0}(\bm{k})$ are given by Eq.~(\ref{eq:ham0}).
Introducing the 3-component vector, $\bm{\mathcal{D}}(\phi)=(\psi/\sqrt{2}$, $D_{z}$, $d_{z}$), the BCS gap equation is given by
\begin{align}
&\mathcal{D}_{i}(\phi)=-\sum_{jk}v_{ik}K_{kj}(\phi)\mathcal{D}_{j}(\phi),
\cr
&K_{ij}(\phi)=\frac{T}{N_{0}}\sum_{\bm{k}n}{\rm Tr}\left[
X^{i\dagger}G^{(0)}(\bm{k},i\omega_{n})X^{j}G^{\rm T}(-\bm{k},-i\omega_{n};\phi)
\right],
\quad
\label{eq:gap}
\end{align}
where
\begin{align}
&G^{}(\bm{k},i\omega_{n};\phi)=\biggl[i\omega_{n} - \mathcal{H}_{\rm BdG}(\bm{k};\phi)\biggr]^{-1}_{11},
\end{align}
is the matrix Matsubara Green's function in the superconducting state (the subscript $11$ represents the upper-left component in Nambu space), and that with the superscript $(0)$ is for the normal state.
Here, $\omega_{n}=(2n+1)\pi T$ is the fermionic Matsubara frequency.
The interaction matrix elements are given in terms of $U$, $J_{0}$, and $J_{2}$ as
\begin{align}
&
v_{11}=\frac{U}{4},
\quad
v_{22}=\frac{1}{4}(U+J_{2}),
\quad
v_{12}=v_{21}=\frac{J_{0}}{2\sqrt{2}},
\cr&
v_{33}=\frac{1}{4}(U-3J_{2}).
\label{v3}
\end{align}

We solve the gap equation with the fixed electron density $n$, which is given by
\begin{align}
n=\frac{T}{N_{0}}\sum_{\bm{k}n}\sum_{m\sigma}[G(\bm{k},i\omega_{n};\phi)]^{\sigma\sigma}_{mm}e^{i\omega_{n}0_{+}}, \label{eq:n}
\end{align}
and the doping rate is related as $x=2-n$. 
 In the actual calculation of Eqs. (\ref{eq:gap}) and (\ref{eq:n}), we analytically sum over the Matsubara frequencies and numerically perform the summation of the wave vector on the uniform $\bm{k}$-meshes.

\textit{Phase diagram |}
We solve the self-consistent gap equation for the fixed interaction strength $U=-0.5$ (attractive) and $J=-U/1.7$ eV (ferromagnetic).
Figure~\ref{fig:1}(b) shows the overall phase diagram in the temperature $T$ and doping $x$.
The arrows in the circles represent the phase relation between $\psi$, $d_{z}$, and $D_{z}$ in the complex plane. 
The TRS breaking phase (TRBP) appears at around $x\sim0.1$.
The inset indicates the change of the relative phase between $\psi$ and $d_{z}$, where the intermediate phase between $0$ and $\pi$ stands for the TRS breaking state.
At this doping rate, the chemical potential touches the lower spin-split band, and the change of the FS topology (Lifshitz transition) occurs as shown in Fig.~\ref{fig:1}(c).  

\begin{figure}[t!]
\begin{center}
\includegraphics[width=7.0cm]{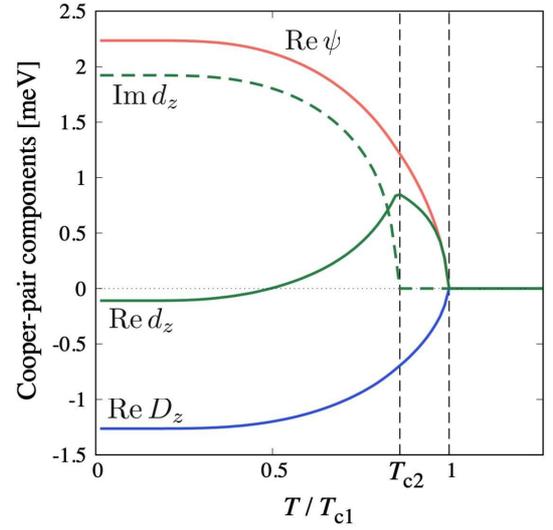}
\caption{
Temperature dependance of the Cooper-pair components at $x=0.103$.
The red, blue, and green lines represent the real part of the $\psi$, $d_{z}$, and $D_{z}$, respectively.
The imaginary part of the $d_{z}$ appears below $T_{\rm c2}$ (the green dashed line), indicating the appearance of the TRS breaking pairing.
}
\label{fig:2}
\end{center}
\vspace{-7mm}
\end{figure}

Thus, the appearance of the TRBP is triggered by the Lifshitz transition. 
Note that in both sides of the Lifshitz transition point, $\psi$ is the predominant component (the $T$ dependence of the Cooper-pair components is shown in Fig.~\ref{fig:2} for the case $x=0.103$), and in the left (right) side the anti-phase (in-phase) between $\psi$ and $d_{z}$ turns out to be energetically stable.
As a result, around the Lifshitz transition the $d_{z}$ changes its sign by passing through the complex plane in order to gain the condensation energy.
This is the reason why TRBP appears near the Lifshitz transition.

It should be emphasized that without the SOC, $d_{z}$ and $\psi$ belong to the different irreducible representation, and they do not hybridize with each other.
Moreover, no spin splitting arises without the SOC, which eliminates the Lifshitz transition itself.

In many cases, the relative phase of $0$ or $\pi$ is usually stabilized for the superconducting state having two non-degenerate components, yielding TRS preserving states~\cite{Y.Tanaka(2),Suhl, Peretti, Kondo}
On the contrary, for more than three components with appropriate interactions between them, a sort of frustration is introduced in their relative phases, and the resulting superconducting state would break TRS.

\textit{Topological classification | }
To elucidate the nature of the TRS breaking state, let us first make symmetry argument.
We focus on the transformation properties of the BdG Hamiltonian (\ref{eq:BdG}) with respect to the time-reversal $\theta$, particle-hole $\mathcal{P}$ and chiral $\mathcal{C}$ symmetry operations, which are represented by
\begin{align}
&
\theta = \left(\tau^{0}/\sqrt{2} + \tau^{x}\right) \otimes (i \sigma_{y}) \, \mathcal{K},
\\
&
\mathcal{P} = \left(\tau^{0}/\sqrt{2} + \tau^{x}\right) \otimes  \rho^{x},
\\
&
\mathcal{C} = \mathcal{P} \theta,
\end{align}
respectively, where $\mathcal{K}$ is the complex conjugation, and the Pauli matrix $\rho^{x}$ acts on the particle-hole Nambu space.
The BdG Hamiltonian (\ref{eq:BdG}) is transformed as
\begin{align}
&
\theta\, \mathcal{H}_{\rm BdG} (\bm{k}; \phi)\, \theta^{-1} = \mathcal{H}_{\rm BdG} (-\bm{k}; -\phi),
\label{tconv}
\\
&
\mathcal{P}\, \mathcal{H}_{\rm BdG} (\bm{k}; \phi)\, \mathcal{P} ^{-1} = - \mathcal{H}_{\rm BdG} (-\bm{k}; \phi),
\label{pconv}
\\
&
\mathcal{C}\, \mathcal{H}_{\rm BdG} (\bm{k}; \phi)\, \mathcal{C} ^{-1} = - \mathcal{H}_{\rm BdG} (\bm{k}; -\phi),
\label{cconv}
\end{align}
and the symmetry classification is summarized in Table~\ref{tab:1}.
From these properties, it is clear that both the time-reversal and chiral symmetries are broken for $\phi\neq 0,\pi$.
According to the topological classification~\cite{A.P.Schnyder,S.Ryu,A.Altland,A.Yu.Kitaev}, our TRS breaking state with $\phi \neq 0$, $\pi$ belongs to the class D, which is the same class as the chiral $p_{x}+ip_{y}$ pairing as was discussed in Sr$_2$RuO$_4$.

\begin{table}[t!]
\begin{center}
\caption{
Symmetry classification~\cite{A.P.Schnyder,S.Ryu,A.Altland,A.Yu.Kitaev} of the BdG Hamiltonian (\ref{eq:BdG}) in terms of the time-reversal $\theta$, particle-hole $\mathcal{P}$, and chiral $\mathcal{C}$ symmetry operations, and the corresponding topological properties (TP).
The values of $\theta^{2}$ and $\mathcal{P}^{2}$ are shown in the second and third columns.
The value $0$ ($1$) indicates the absence (presence) of $\mathcal{C}$ symmetry.
The last two columns represent the conditions for finite orbital and spin angular-momentum distribution, where the subscripts ``s'' and ``a'' indicate symmetric and antisymmetric parts of the momentum-space distribution, respectively.
}
\label{tab:1}
\begin{tabular}{cccccccc}
\hline\hline
Phase & $\theta$ & $\mathcal{P}$ & $\mathcal{C}$ & Class & TP ($d=2$) & $l^{z}_{\rm s}$, $s^{z}_{\rm s}$ & $l^{z}_{\rm a}$, $s^{z}_{\rm a}$ \\
\hline
SC ($\phi \neq 0$) & $0$ & $+1$ & $0$ & D & $\mathbb{Z}$ & $\bigcirc$ & $\bigcirc$ \\
SC ($\phi =     0$)  & $-1$ & $+1$ & $1$ & DIII & $\mathbb{Z}_{2}$ & $\times$ & $\bigcirc$ \\
Normal & $-1$ & $+1$ & $1$ & --- & --- & $\times$ & $\bigcirc$ \\
\hline\hline
\end{tabular}
\end{center}
\vspace{-7mm}
\end{table}

\textit{Momentum distribution of the angular momentum |}
Next, we discuss the momentum-space distributions of the orbital and spin angular momentum, which characterize the TRS breaking state.
The distribution is defined as the thermodynamic average at each temperature,
\begin{align}
M(\bm{k};\phi) = \sum_{\alpha\beta}
M_{\alpha \beta}\Braket{c^{\dagger}_{\bm{k} \alpha} c^{}_{\bm{k} \beta}} = - \sum_{\alpha\beta}M^{\mathrm{T}}
_{\alpha \beta} \Braket{c^{}_{\bm{k} \alpha} c^{\dagger}_{\bm{k} \beta}},
\end{align}
where $M$ is a traceless matrix representing either the orbital $l^{z}$ or spin $s^{z}$ operators.
The other components vanish due to the $xy$-plane mirror symmetry.
Since the antisymmetric spin splitting always exists even in the normal state, we discuss separately the symmetric and antisymmetric parts of the momentum distribution, \textit{i.e.,} $M_{\rm s,\,a}(\bm{k})=[M(\bm{k})\pm M(-\bm{k})]/2$.

\begin{figure}[t!]
\begin{center}
\includegraphics[width=7.0cm]{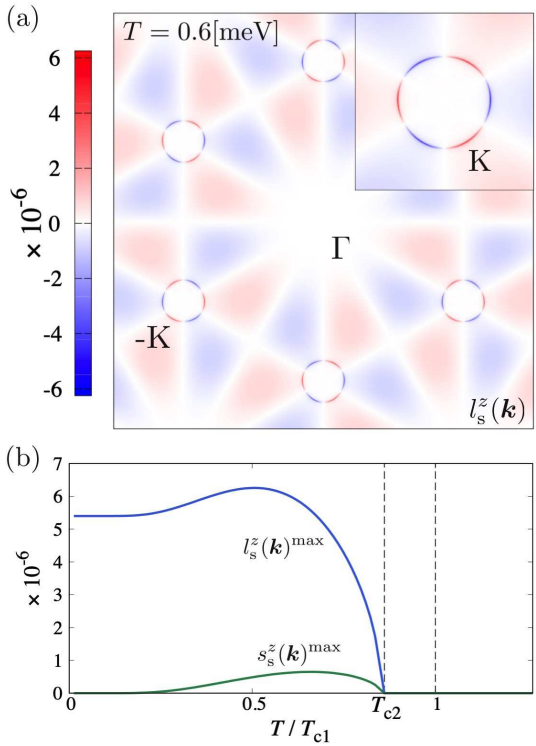}
\caption{
(a) Symmetric part of the momentum distribution of the orbital angular momentum $l^{z}_{\rm s} (\bm{k};\phi > 0)$ in the TRS breaking phase at $T = 0.6$ meV.
The numerically obtained $l^{z}_{\rm s}(\bm{k};\phi > 0)$ is indicated by the solid lines on the Fermi surfaces around K and K$'$ points, while the background contour map represents the A$'_{2}$-symmetry form factor $f(\bm{k})$ given by Eq.~(\ref{eq:formf}). 
(b) The temperature dependance of the symmetric part of the spin and orbital angular momentum distribution $s^{z}_{s} (\bm{k}; \phi > 0)$ and $l^{z}_{s} (\bm{k}; \phi > 0)$; the maximum values are taken from the  momentum-space distributions.
}
\label{fig:3}
\end{center}
\vspace{-7mm}
\end{figure}

According to Eqs.~(\ref{tconv})--(\ref{cconv}), it is shown that the $M_{\rm s,\,a}(\bm{k})$ satisfies the following relations, 
\begin{align}
&M_{\rm s,\, a} (\bm{k}; \phi)=\mp M_{\rm s,\, a} (\bm{k}; -\phi) ,\\
&M_{\rm s,\, a} (\bm{k}; \phi)= \pm M_{\rm s,\, a} (-\bm{k}; \phi), \\
&M_{\rm s,\, a} (\bm{k}; \phi)=  -M_{\rm s,\, a} (-\bm{k}; -\phi).
\end{align}
Therefore, the symmetric parts $s^{z}_{\rm s}(\bm{k}; \phi)$ and $l^{z}_{\rm s}(\bm{k}; \phi)$ can be finite only in the TRS breaking phase.
The conditions for finite $M_{\rm s,a}$ are summarized in Table~\ref{tab:1}.

We further discuss the momentum dependences of the symmetric parts $s^{z}_{\rm s}(\bm{k};\phi)$ and $l_{\rm s}^{z}(\bm{k};\phi)$.
Since the TRS breaking state is characterized by the gap function in fully symmetric A$'_{1}$ irreducible representation, the BCS mean field should preserve the point-group symmetry.
When we decouple the mean-field term as $\mathcal{H}_{\rm MF}=f(\bm{k})(\alpha l^{z}+\beta s^{z})$, the form factor $f(\bm{k})$ must have A$'_{2}$ symmetry since $l^{z}$ and $\sigma^{z}$ belong to A$_{2}'$ symmetry. 
The lowest-order form factor in A$_{2}'$ is antisymmetric with respect to $\bm{k}$, giving rise to the antisymmetric spin splitting.
The next lowest-order form factor is symmetric and is given by
\begin{multline}
f(\bm{k}) = 2 \left[ \sin \left(\frac{3 \sqrt{3} k_{x}}{2}\right) \sin \left(\frac{k_y}{2}\right) - \sin \left(\sqrt{3} k_{x}\right) \sin \left(2 k_{y} \right)\right.
\\
+  \left.\sin \left(\frac{\sqrt{3} k_{x}}{2}\right) \sin \left(\frac{5 k_y}{2}\right)\right].
\label{eq:formf}
\end{multline}
Figure~\ref{fig:3}(a) shows the calculated momentum-space distribution (indicated by the solid lines) of $l^{z}_{\rm s} (\bm{k}; \phi > 0)$ in the TRS breaking phase at $x=0.103$ and $T = 0.6$ meV.
The finite distribution appears on the FS around K and K$'$ points, and exhibits three-fold symmetry reflecting the symmetric A$_{2}'$ form factor.
The background contour map represents the form factor $f(\bm{k})$ in Eq.~(\ref{eq:formf}), which is consistent with the calculated $l_{\rm s}^{z}(\bm{k})$ appeared on the FS.

The $T$ dependences of the maximum values of the distributions are shown in Fig.~\ref{fig:3}(b).
The finite distributions appear at the onset of the TRBP.
Note that the distribution of the orbital angular momentum saturates to the finite value at $T=0$, while that of the spin angular momentum decreases exponentially toward $T=0$.
This is because the fact that in the spin-triplet $d_{z}$ state the spins coming from condensation pairs lie in the $xy$ plane, and the $z$-component spins of the quasiparticle excitations only contribute to the finite angular momentum distribution.

\textit{Summary |}
In the present work, we have revealed the nature of
the $s$-wave TRS breaking superconductivity, which emerges in the realistic three-band Hubbard-like model for the hole-doped monolayer MoS$_{2}$. 
It is found that this unusual TRS breaking phase is triggered by the Lifshitz transition, where the three distinct gap components, $\psi$, $d_{z}$, and $D_{z}$, compete with each other, resulting in the complex gap functions in the subdominant components.
The competition of the gap functions and the Lifshitz transition itself are caused by the atomic SOC in the noncentrosymmetric crystal structure.
The TRS breaking pairing state is characterized by D class in the topological classification, and finite momentum distributions of the orbital and spin angular momentum with three-fold rotational symmetry in A$_{2}'$ irreducible representation on the FS pockets around K and K$'$ points.
This intriguing TRS breaking state provides us with an archetypal example based on the realistic model, which is a representative case of the spin-orbital-coupled metals in noncentrosymmetric crystal structure.

\begin{acknowledgment}
The authors would like to thank S. Sumita, P.M.R. Brydon, S. Hayami, and K. Miyake for fruitful discussions.
This work was supported by JSPS KAKENHI Grants Number JP15H05885 (J-Physics).
\end{acknowledgment}

\end{document}